\documentclass[a4paper,american]{scrartcl}

\usepackage{tikz-cd}
\usepackage{verbatim}
\usepackage{enumerate}
\usepackage{lmodern}
\usepackage[utf8]{inputenc}
\usepackage[T1]{fontenc}
\usepackage{color}
\usepackage{todonotes}
\usepackage{babel}
\usepackage{amsfonts, amsmath, amsthm, amssymb}
\usepackage{setspace}
\usepackage[authoryear]{natbib}
\usepackage[font=small]{quoting}
\usepackage{latexsym}
\usepackage[bookmarks=false,pdfborder={0 0 0},colorlinks=false]{hyperref}
\usepackage{dsfont}
\usepackage{xcolor}
\makeatletter

\newcommand{\IR}{\mathbb R}
\newcommand{\IP}{\mathbb P}

\theoremstyle{plain}

\theoremstyle{definition}

\title{Position Measurements and the Empirical Status of Particles in Bohmian Mechanics }
\author{ Dustin Lazarovici\thanks{Dustin.Lazarovici@unil.ch}}

\begin{document}
	\maketitle
	
\begin{abstract} \noindent The paper addresses the debate about the empirical status of particles versus wave functions in Bohmian quantum mechanics. It thereby clarifies questions and misconceptions about the role of the particles in the measurement process, the (un)reliability of position measurements (``surrealistic trajectories''), and the limited empirical access to particle positions (``absolute uncertainty''). Taking the ontological commitment of Bohmian mechanics seriously, all relevant empirical results follow from an analysis of the theory in terms of particle motions. Finally, we address the question, why particle motions rather than patterns in the wave function would be the supervenience base of conscious experience.
\end{abstract}

\section{Introduction}

Bohmian mechanics is a quantum theory based on a primitive ontology of particles and two precise mathematical equations defining their dynamics. These are the Schrödinger equation
\begin{equation}\label{eq:schroedinger}
	i\hbar \partial_t \psi_t
	=
	H\psi_t,  
\end{equation}
for the wave function, and the guiding equation
 \begin{align}
 	\label{eq:BM-vel}
 	\dot{X}_k = v^\psi_{k,t}(X) := 
 	\frac{\hbar}{m_k}\mathrm{Im} \frac{\nabla_k\psi_t(X)}{\psi_t(X)},
 \end{align}
 in which the wave function enters to determine a velocity field for $N$ particles with positions $X=(X_1,\ldots,X_N)\in \IR^{3N}$.  On the fundamental level, there is only one wave function, the universal wave function, guiding the motion of all the particles together. In many relevant situations, however, subsystems allow for an autonomous description in terms of an \emph{effective wave function} determined by the universal wave function and the actual positions of particles outside the subsystem. It can then be shown that Born's rule, applied to effective wave functions, describes the \emph{typical} distribution of particle positions in an ensemble of identically prepared subsystems \citep[ch. 2]{durr.etal2013a}. 
 
 With this \emph{quantum equilibrium hypothesis}, the Bohmian theory reproduces the statistical predictions of standard quantum mechanics (whenever the latter are  well-defined). It does so by making correct statistical predictions about the outcome of measurement experiments as recorded in the spatial configuration of whatever plays the role of a ``measurement device'' \citep[ch. 3]{durr.etal2013a}.  
 
 While Bohmians generally insist that the empirical content of the theory is exhausted by its predictions about particle motions, critics have questioned the empirical status of the particles, usually advocating for a priority of the wave function when it comes to relating the theory to observation (e.g. \cite{zeh1999, bedard1999, brown.wallace2005, gao2019a}). On this basis, it has even been argued that Bohmian mechanics doesn't solve the quantum measurement problem \citep{stone1994, gao2019}, or that it solves the measurement problem only by being a Many-Worlds theory in denial \citep{deutsch1996}. The misleading terminology of ``hidden variables'' has probably done its part to stir the debate about just how hidden the Bohmian particles actually are. 
 
 I  will argue that these criticisms are based on misconceptions of the Bohmian theory and the role of particles vis-a-vis wave functions in it. To the extent that valid questions have been raised -- in particular about the empirical accessibility of particle positions -- they are questions that can be answered. To this end, I will first provide a brief review of the Bohmian description of the measurement process. Section 3 will clarify the status of particles and wave functions in Bohmian mechanics and address various worries about the empirical (in)accessibility of particle positions. In Section 4, I will (reluctantly) address the issue of conscious experience and why, assuming a functionalist theory of mind, mental states would be realized by the particles rather than the wave function. I end with a short ``dialogue'' in Section 5, trying to put the discussion into a broader perspective.

 \section{The measurement process in Bohmian mechanics}
 
 A prototypical measurement in Bohmian mechanics is an interaction between a system $S$ and a measurement device $D$ resulting in one of several macroscopically discernible configurations of $D$ (``pointer positions'') which are correlated with certain possible quantum states of $S$. Schematically, the interaction between the measured system and measurement device is such that, under the Schrödinger evolution,
 \begin{equation} \label{Pfeil1}
 \varphi_i\Phi_0\stackrel{\mbox{\footnotesize
 		Schrödinger evolution\index{Schrödinger! Entwicklung}}}{\longrightarrow}\varphi_i\Phi_i\,\,,
 \end{equation}
 where the wave function $\Phi_0$ is concentrated on pointer configurations corresponding to the ``ready state'' of the measurement device, and $\Phi_i$ are concentrated on configurations indicating a particular measurement result, e.g., by a pointer pointing to a particular value on a scale, a point-like region of a detector screen being darkened, a detector clicking or not clicking, etc. The Schrödinger time evolution is linear, so that a superposition
 \[\varphi=c_1\varphi_1+c_2\varphi_2,\qquad
 c_1,c_2\in\mathbb{C},\qquad |c_1|^2+|c_2|^2=1,\] leads to
 \begin{equation} \label{Pfeil2}
 \varphi\Phi_0=(c_1\varphi_1+c_2\varphi_2)\Phi_0
 \stackrel{\mbox{\footnotesize
 		Schrödinger evolution\index{Schrödinger! Entwicklung}}}{\longrightarrow}
 c_1\varphi_1\Phi_1+c_2\varphi_2\Phi_2.
 \end{equation}
 At this point, standard quantum mechanics is hit by the measurement problem \citep{maudlin1995e}.  In Bohmian mechanics, however, the system is described not only by the wave function but also by the actual spatial configuration $(X,Y) \in \IR^k \times \IR^m$ of measured system and measurement device, given by the positions of their constituent particles. It thus has a well-defined configuration at all times, regardless of whether or not its wave function is in a superposition. 
 
 \begin{figure}[ht]
 	\begin{center}
 		\includegraphics[width=\textwidth]{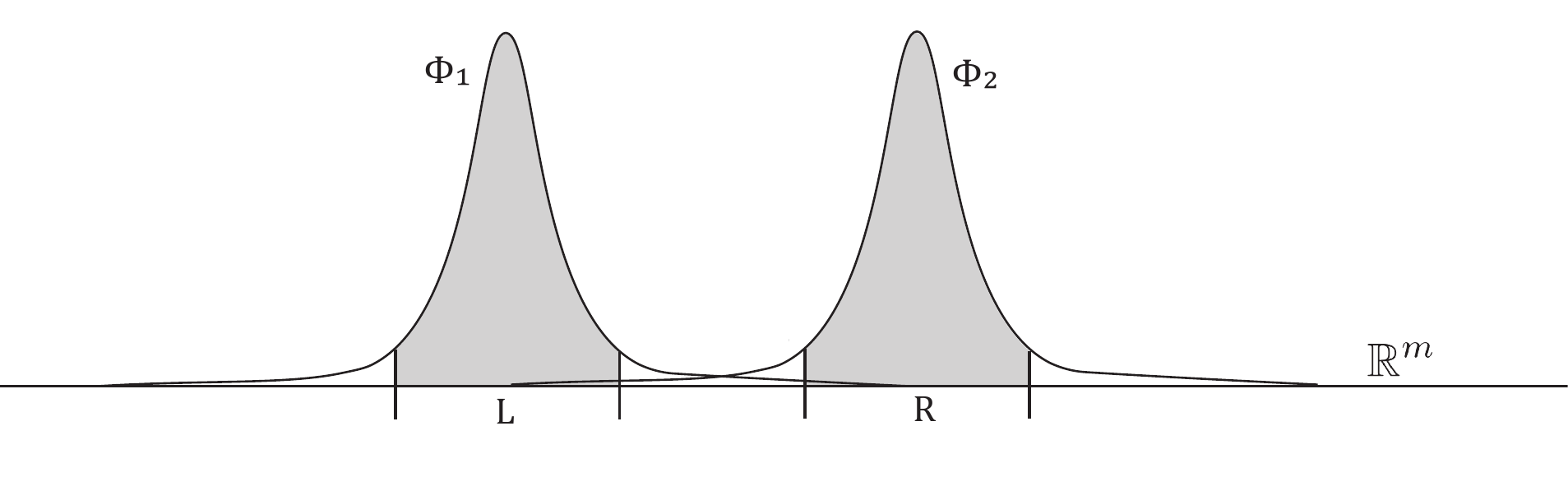}
 		\caption{Sketch of the pointer wave functions on configuration space.}\label{fig:pointerstates}
 	\end{center}
 \end{figure}

 For illustrative purposes, we assume that $\Phi_1$ is concentrated on a region $\mathrm{L}\subset \IR^m$ of the configuration space of $D$ corresponding to pointer-configurations pointing to the left, while $\Phi_2$ is concentrated on a region $\mathrm{R}\subset \IR^m$ corresponding to pointer-configurations pointing to the right. Obviously, the two regions are disjoint, i.e. $\mathrm{L}\cap \mathrm{R}=\emptyset$. By assumption, $\Phi_1$ and $\Phi_2$ are well localized in the respective regions (otherwise, the measurement device is no good), i.e., almost zero outside (see Fig. 1). In particular, we have
 \begin{subequations}
 	\begin{align}\label{L1}
 	\int_{\mathrm{L}} |\Phi_1|^2 \;  \mathrm{d}^my \approx 1, \;\;\;
 	\int_{\mathrm{L}} |\Phi_2|^2 \; \mathrm{d}^my \approx 0\\\label{R0}
 	\int_{\mathrm{R}} |\Phi_1|^2 \; \mathrm{d}^my \approx 0, \;\;\;
 	\int_{\mathrm{R}} |\Phi_2|^2 \; \mathrm{d}^my \approx 1.
 	\end{align}
 \end{subequations}

 Now, according to Bohmian mechanics, the probability of the pointer \emph{actually} pointing to the left is: 
 \begin{equation}\begin{split}\label{pointerintegral}
 \IP(Y \in \mathrm{L}) &= \int_{\IR^k \times \mathrm{L}}|c_1\varphi_1\Phi_1+ c_2\varphi_2\Phi_2|^2 \, \mathrm{d}^kx\,\mathrm{d}^my \\
 & =  |c_1|^2 \int_{\IR^k   \times \mathrm{L}}|\varphi_1\Phi_1|^2\mathrm{d}^kx \,\mathrm{d}^my\\& +
 |c_2|^2\int_{\IR^k  \times \mathrm{L}} |\varphi_2\Phi_2|^2\mathrm{d}^kx \,\mathrm{d}^my\\ 
 &+2 \,\mathrm{Re} \Bigl( c_1c_2\int_{\IR^k  \times \mathrm{L}}(\varphi_1\Phi_1)^*
 \varphi_2\Phi_2 \mathrm{d}^kx \,\mathrm{d}^my \Bigr) \approx|c_1|^2.\end{split}
 \end{equation}
 The final approximation follows from eq. \eqref{L1} (together with the Cauchy-Schwarz inequality $\left\lvert \int_{L} \Phi_1^* \Phi_2 \right\rvert \leq  \sqrt{\int_{L} |\Phi_1|^2}\sqrt{\int_{L} |\Phi_2|^2}$ ). Similarly, the probability of the pointer pointing to the right is $\IP(Y \in \mathrm{R})\approx |c_2|^2$. If $\varphi_1$ and $\varphi_2$ are eigenstates of some quantum observable, $|c_1|^2$ and $|c_2|^2$ are the statistical predictions of standard quantum mechanics for an ideal measurement. (The better the pointer states $\Phi_1$ and $\Phi_2$ are localized in disjoint regions of configuration space, the closer the measurement is to ``ideal''. )

 Moreover, after the measurement (assuming it was not destructive), the measured system $S$ will be guided by the wave function 
 $\varphi_1(x)\Phi_1(Y) + \varphi_2(x)\Phi_2(Y)$. If the pointer actually points left (let's say), i.e. $Y \in \mathrm{L}$, we have $\Phi_2(Y) \approx 0$ and hence (after normalization) the effective wave function $\varphi_1$ describing the system $S$ at the end of the measurement. In this way -- that depends crucially on actual particle positions -- Bohmian mechanics vindicates the postulate of textbook quantum mechanics that a measurement collapses the wave function of the measured system such that the previous outcome will be reproduced by a repeated measurement. It does not, however, vindicate the (bad) idea that the state $\varphi_1$ or $\varphi_2$ corresponds to some pre-existing property of the system (``observable value'') that the measurement merely reveals (cf. \cite{lazarovici.etal2018}). 
 Bohmian particles have a position and nothing else, while the physical content of the wave function is understood through its role for the dynamical and statistical description of the particles.




\section{The epistemic status of particles}
Despite this central role of point particles in Bohmian mechanics -- or maybe because of it -- there has been a lot of debate about their empirical status.  

Some authors have suggested that Bohmian mechanics includes -- or should include -- a postulate stating that measurement results are instantiated in particle positions, or that observations ``supervene'' on particle positions (rather than the wave function), or something like that (see e.g. \cite{naaman-marom.etal2012}). Such a postulate is neither helpful nor necessary, as I hope to clarify with this paper. In fact, Bohmians generally insist (as did John \citet[ch. 23]{bell2004}) that it is a bad idea to include postulates about ``observation'' or ``measurements'' in any physical theory since those are much too vague and physically complex notions.

Other authors suggest that ``measurement results'' in Bohmian mechanics correspond first and foremost to certain wave functions, while the role of the particle configuration is merely to ``pick out'' one part of a (decoherent) superposition as the actual result. In particular, \cite{brown.wallace2005}  claim to identify such a ``Result Assumption'' in the second part of David Bohm's 1952 paper.\footnote{\cite{bohm1952a} writes: ``[T]he packet entered by the apparatus  variable $y$ determines the actual result of the measurement, which the observer will obtain when he looks at the apparatus.'' (p. 182)} I lack the historical competence to provide a thorough exegesis of Bohm's original work. I believe that Brown and Wallace are reading too much into an innocuous statement, but can't rule out the possibility that Bohm had not yet appreciated the implications of his theory in full. What I can unequivocally say is that such a ``Result Assumption'' plays no role in the modern understanding of Bohmian mechanics (that has been further developed by Bell, and Dürr, Goldstein, Zanghì, among others). Indeed, it would be a rather unproductive assumption to make since it leaves open the critical question, how and why and in what sense a particular wave function is supposed to ``correspond to a measurement result'' -- or any concrete physical fact at all. 

Unsurprisingly, though, this $\psi$-centric reading has resonated in particular with modern Everettians who are committed to the view that objects and events in physical space (like measurement devices indicating a measurement result) can be recovered by some sort of functional analysis in terms of internal degrees of freedom of the wave function or quantum state. This, however, is \emph{not} how the Bohmian theory relates to the physical world, and there are legitimate questions as to whether the procedure can succeed in general (see e.g. \cite{monton2006, maudlin2010}; I will express some of my own concerns in the course of this paper). 

What Bohmian mechanics makes is an ontological commitment to particles. They are the local beables \citep[ch. 7]{bell2004} or primitive ontology \citep{allori.etal2014}, what the theory postulates as the basic constituents of matter. The role of the wave function is first and foremost to determine the motion of particles and also (though this is a theorem rather than an additional postulate) to describe their statistical distributions. All our analyses of the theory are then consistent with the particles forming stable configurations that move and behave, qualitatively and quantitatively, like the tables, cats, measurement devices, etc. that we observe in the world. This is why the theory is empirically adequate. In particular, the way in which the particle ontology solves the measurement problem is not just by picking out certain parts or branches of the wave function as guiding but by releasing the wave function from the undue burden of representing matter in the first place. 


A point that Bohmians \emph{do} repeatedly and emphatically insist on, is that making correct predictions about the spatio-temporal configuration of matter -- including pointer positions, display readings, or whatever else is used to ``record'' the outcome of ``measurements'' -- is sufficient for the empirical adequacy of a physical theory (cf. \citet[p. 166]{bell2004}). But this is a claim about physics in general, not an additional postulate about Bohmian measurements in particular. It is unfortunate since potentially misleading that some authors (e.g. \cite{gao2019}) mistake it for the latter. 

As a nod to the neo-Everettians (and other wave function monists), it is worth pointing out that Bohmians are also ``macro-object functionalists'' (\cite{lewis2007}) in the sense that functionalist arguments are relevant to locating macroscopic objects in the particle trajectories. However, while I understand how things moving and interacting in physical space can be functionalized in terms of other things moving and interacting in physical space, it is unintelligible to me how things moving and interacting in physical space could be functionalized in terms of degrees of freedom of the wave function which (no matter how you want to think about it) are not things moving and interacting in physical space. I will return to this issue in Section 4.

\subsection{Position Measurements}

Some sceptics now say that this is all well and good, Bohmian mechanics may predict that particles can form cats and tables and measurement devices that have a definite configuration at all times, but there is no good reason to believe that when we \emph{look} where a table is or whether the pointer points left or right, we will see them in the position that the theory predicts for the particles.

The intuition behind this worry seems to be that observations are physical interactions and that these interactions are first and foremost described by the Schrödinger equation for wave functions which makes no reference to Bohmian particle positions. Hence, it may seem like observations are determined by the wave function after all, while the particles are somehow epiphenomenal.

This reasoning is not correct, but since an observation is indeed a physical interaction, the question here is ultimately a physical one, so let's see what the theory actually predicts. We recall the measurement procedure described in Section 2 with the final wave function of system and apparatus given by the right-hand-side of eq. \eqref{Pfeil2}. Now we go one step further and consider a ``measurement of the pointer position'' by another system $C$ (we assume that the measurement device $D$ was perfectly isolated up to this point, so there is no environmental decoherence). We may think of an ``observer'' looking at the measurement device, resulting, ultimately, in a certain particle configuration of her brain, though I prefer a camera or some other system under no suspicion of consciousness (we will return to the issue of conscious experience in Section 4). In any case, the spatial resolution of such an observation can very well be finer than the spread of the ``pointer states'' $\Phi_i$ (thus corresponding to a Schrödinger evolution $
\Phi_i \longrightarrow \sum_j \Phi_{ij} \Psi_{j},
$
where $\sum_j \Phi_{ij} = \Phi_{i}$ and the $\Psi_{j}$ are the ``record states'' of $C$.) However, we shall consider the simplest case in which the  measurement interaction leads to a final wave function of the form
\begin{equation}
\, c_1\varphi_1\Phi_1\Psi_1+c_2\varphi_2\Phi_2\Psi_2, \,
\end{equation}
where $\Psi_1$ is concentrated on a region $\mathcal{L}$ of the configuration space of $C$ corresponding to the camera recording a pointer pointing left, and $\Psi_2$ is concentrated on a region $\mathcal{R}$ corresponding to the camera recording a pointer pointing right. 

So what is the probability that the pointer actually points to the left, i.e. $Y \in \mathrm{L}$, while the camera records a pointer pointing right, i.e. $Z \in \mathcal{R}$?  We find
\begin{equation}\label{pointerintegral}
\IP(Y \in \mathrm{L}, Z \in \mathcal{R}) = \int_{\IR^k \times \mathrm{L}\times \mathcal{R}} |c_1\varphi_1\Phi_1\Psi_1+ c_2\varphi_2\Phi_2\Psi_2|^2 \, \mathrm{d}^kx\, \mathrm{d}^my\, \mathrm{d}^nz \approx 0,
\end{equation}
since $\Phi_2$ is zero (or nearly so) on $\mathrm{L}$, while $\Psi_1$ is zero (or nearly so) on $\mathcal{R}$, hence both $\Phi_1\Psi_1$ and $\Phi_2\Psi_2$ are just about zero  on $\mathrm{L}\times \mathcal{R}$. Simply put: if you look where the pointer is, you will typically see the pointer where it is.

\begin{figure}[ht]
	\begin{center}\label{fig:confspace}
		\includegraphics[width=\textwidth]{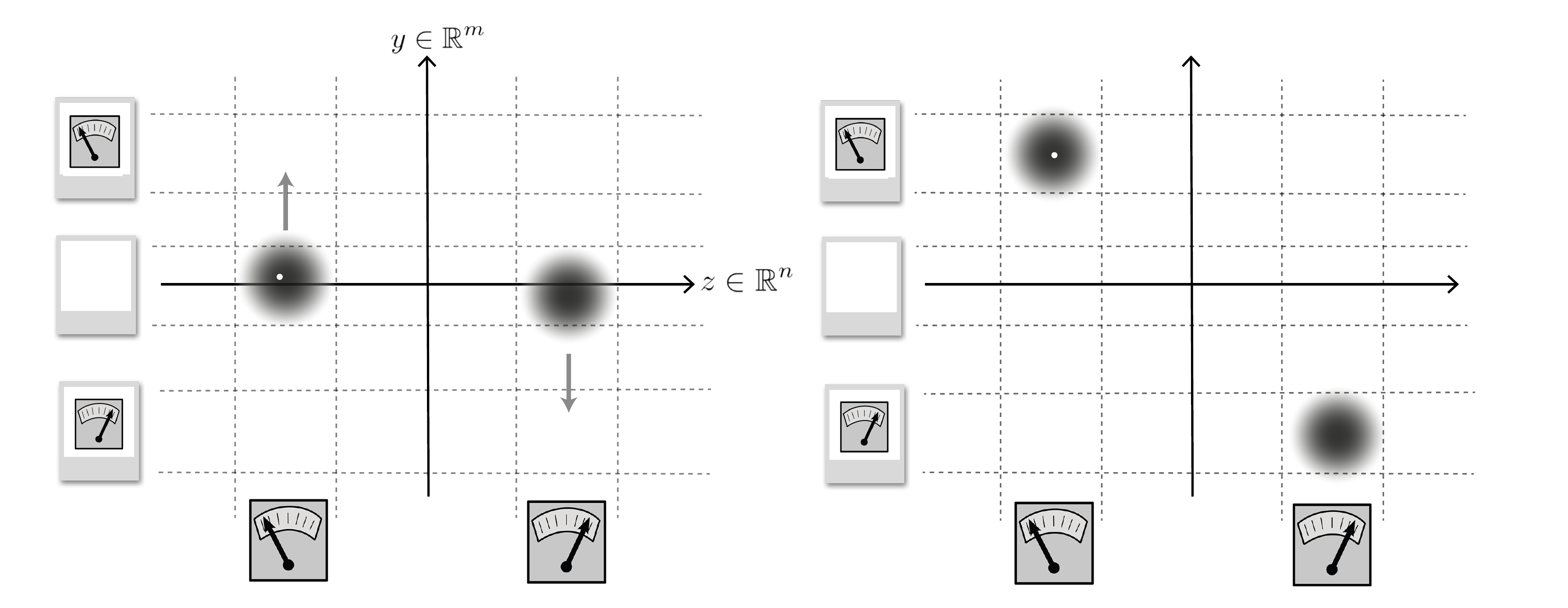}
		\caption{Sketch of position measurement in configuration space. The dot indicates the actual configuration of the system.}
	\end{center}
\end{figure}

How does this result square with the argument that particle positions do not matter because interactions are described by the wave function and its Schrödinger evolution? Well, as I said, the argument is not correct (see also \cite{maudlin1995d}). It neglects the fact that in the interaction between the systems $D$ and $C$, the particle configuration of $D$ is essential to determining which part of the wave function guides the configuration of $C$. It is instructive to consider an intermediate stage of the measurement interaction
\begin{equation}
(\Phi_L + \Phi_R)\Psi_0 \longrightarrow \Phi_L\Psi_{\vartriangleleft} + \Phi_R\Psi_{\vartriangleright} \longrightarrow \Phi_1\Psi_1+ \Phi_2\Psi_2
\end{equation}
in which the wave packets $\Psi_{\vartriangleleft}$ and  $\Psi_{\vartriangleright}$ are just beginning to separate in the configuration space of $C$ and propagate towards the regions $\mathcal{L}$ and $\mathcal{R}$, respecticely. Note that in the full configuration space of $D+C$, however, the entangled wave function $\Phi_L\Psi_{\vartriangleleft} + \Phi_R\Psi_{\vartriangleright}$ is already well-separated (decohered) along the $y$-coordinates (Fig. 2).   
Now, acccording to the guiding equation \eqref{eq:BM-vel}, the velocity of the $Z$-variables  is

\begin{equation}
\dot{Z} \propto \mathrm{Im} \frac{\Phi_L(Y)\nabla_z\Psi_{\vartriangleleft}(Z) + \Phi_R(Y)\nabla_z\Psi_{\vartriangleright}(Z)}{\Phi_L(Y)\Psi_{\vartriangleleft}(Z) + \Phi_R(Y)\Psi_{\vartriangleright}(Z)}.
\end{equation}

Hence, if the pointer is actually left, i.e. $Y \in \mathrm{L}$, we have $\Phi_R(Y) \approx 0$ and thus $\dot{Z} \approx \mathrm{Im} \frac{\nabla_z\Psi_{\vartriangleleft}(Z)}{\Psi_{\vartriangleleft}(Z)}$, so that the  configuration $Z$ is effectively guided by the wave packet $\Psi_{\vartriangleleft}$ that moves towards $\mathcal{L}$ (i.e. towards configurations in which the photography shows the pointer pointing left). Analogously, if  the observed system is actually right, i.e. $Y \in \mathrm{R}$  the configuration $Z$ is effectively guided by the wave packet $\Psi_{\vartriangleright}$ that moves towards $\mathcal{R}$ (i.e. configurations in which the photography shows the pointer pointing to the right). Hence, the idea that the particles are causally inert, an ``idle wheel'', is clearly wrong. Indeed, it is misleading to say that interactions in Bohmian mechanics are described only by the wave function and the Schrödinger equation; the wave function rather mediates interactions between particles via the guiding law \eqref{eq:BM-vel}.

\subsection{Atypical outcomes}
If we return to the probability estimate, eq. \eqref{pointerintegral}, and suppose that the wave packets $\Phi_i$ or $\Psi_i$ have long ``tails'', $\IP(Y \in \mathrm{L}, Z \in \mathcal{R})$  may indeed not be exactly zero but only nearly so (as indicated by the $\approx$ sign). Hence, there would be a very small, yet non-zero probability that the pointer configuration points to the left (at least for a short period of time), while the camera -- or ``observer'' -- sees a pointer pointing to the right. Realistically, this probability will be so small as to be practically negligible, but the atypical outcome is still \emph{possible} according to the theory. Would this mean that the Bohmian particle configuration $Y$ does not correspond to the ``real'' pointer position? No, it means precisely what the theory says, namely that there is an extremely small, yet non-zero probability that the pointer points left, while the camera records a pointer pointing right. 

And this shouldn't be all that surprising upon reflection. Also according to electrodynamics, it is possible, yet extremely unlikely, that I see the moon to my right while it is actually to my left -- because what I see is a very special, random fluctuation in the electromagnetic field. It is also possible, yet extremely unlikely, that I hold a thermometer (or my finger) in hot water but register a very low temperature because all the fast particles happen to stay away from it.

\emph{Atypicality} can always undermine the reliability of observations; consequently, any inference from empirical evidence has to rely on the assumption that the evidence has not been produced by an atypical or very-low-probability event. This is an important insight about physics in general, not a mystery of Bohmian mechanics or quantum mechanics in particular. 



\subsection{``Position measurements'' that do not measure positions}

There are also special measurement procedures in which the relevant ``record states''  $\Psi_1$ and $\Psi_2$ in eq. \eqref{pointerintegral} would have a big overlap in the configuration space of $C$.  These include, in particular, so-called \emph{weak measurements} but also interactions that lead, for instance, to a spin-flip or the excitation of an atom, so that $\Psi_1$ and $\Psi_2$ are orthogonal in Hilbert space but not separated in configuration space. (This cannot be directly observed but the ``read out'' that manifests in particle configurations can be delayed.) From the same equation, it is evident that such procedures will not reliably reveal the actual particle positions (\cite{aharonov.vaidman1996, naaman-marom.etal2012}). There are even interferometer experiments, in which the naive reading of a detector is systematically wrong about the path of a particle, in which a spin flip (let's say) is always produced by a nonlocal effect rather than a Bohmian trajectory passing nearby (nevertheless, the measurement statistics are always correctly predicted by Bohmian mechanics). This has given rise to the catchy accusation that Bohmian mechanics predicts ``surrealistic trajectories'' \citep{englert.etal2014}. In practice, decoherence prevents such situations for macroscopic systems, but as \cite{gisin2018} rightly points out, there is nothing in the Bohmian theory that makes it \emph{in principle} impossible to perform such an experiment with elephants. This is supposed to sound bad. However, stars are even bigger than elephants and General Relativity tells us that they are not always where we see them (literally). As Einstein reminded the young \citet[p. 80]{heisenberg2012}, it is always the physical theory that has to tell us what can be measured and how, i.e., which physical events are correlated in a way that allows us to infer one from the other. Bohmian mechanics tells us that certain measurement procedures (which are much less trivial than just ``looking'') are not reliable ways to detect the position of a particle or an elephant. Of course, concluding from this that we cannot trust observations of Bohmian particles in general, is to commit a similar mistake as the American president when he says that ``you literally can't see'' the F-35 stealth fighter.

\cite{gisin2018} summarizes the situation correctly by saying that not all measurements which are called ``position measurements'' in standard quantum mechanics are actually position measurements in Bohmian mechanics.\footnote{Another instructive example for this fact was already provided in \citet[sec. 7.5]{durr.etal2004}.} Again, this is probably meant to sound bad (for Bohmian mechanics). But what, in fact, is the justification for calling these (or any) experimental procedures ``position measurements'' in standard quantum mechanics? Is it merely because their statistics can be described by some sort of ``position operator''? This is not a physical account of why and how the detector events in question should be systematically correlated with the position of anything. Orthodox quantum mechanics is unable to provide such an account. In fact, it doesn't even contain localized objects with definite positions, leading to the more basic question, what ``position measurements'' are supposed to measure in the first place.


\subsection{Absolute Uncertainty}

An unfortunate source of confusion about the empirical status of particle positions in Bohmian mechnaics is the theorem of  \emph{absolute uncertainty} \citep[ch. 2]{durr.etal2013a}. This theorem states that if the effective wave function of a subsystem $S$ is $\varphi$, an external observer cannot have more information about the particle configuration of that system than provided by the $\lvert \varphi \rvert^2$-distribution. (``Information'' here just refers to a correlation between the configuration of $S$ and the configuration of some other system -- e.g. a brain -- that constitutes a ``record''.)
\citet[p. 757]{lewis2007} then objects that 
\begin{quote}
``this can't be exactly right; the wavefunction,
after all, doesn't determine a unique result for a measurement. So Bohmians note that since an observer can know which wavepacket contains
the particles, the lower bound on the accuracy with which the particle
configuration can be known is actually the squared amplitude of the
occupied wave packet.''\end{quote}

\noindent The theorem is exactly right (it's a theorem, after all). What Lewis seems to forget is that in order to know the actual measurement result, an observer has to look at (interact with) the measurement apparatus. This will effectively collapse the apparatus wave function into an ``occupied'' wave packet consistent with that measurement result and the observer's knowledge of it.\footnote{For another version of this misunderstanding, see \citet[footnote 1]{gao2019}).} 

To counter further misunderstandings, here are some things the theorem \emph{doesn't} imply: 

\begin{enumerate}[i.]

\item Absolute uncertainty doesn't prevent us from determining particle positions to \emph{arbitrary} precision (again, keeping in mind that whatever procedure we use to localize the particle positions can also localize their wave function).  Note that while one usually states the reverse implication, we could just as well say that our knowledge of the particle positions puts a limit on the spread of their wave function. 

To measure a trajectory is, evidently, just to measure the position at different times, though one then has to keep in mind that since the measurement procedure can change (effectively collapse) the effective wave function, it can also significantly change the trajectory, in particular for microscopic systems. 

\item Absolute uncertainty doesn't prevent us from inferring additional information about past trajectories or particle positions. For instance, in the double slit experiment (assuming a suitably symmetric setup) we know on theoretical grounds that particles hitting the screen 
above/below the symmetry axis have passed through the upper/lower slit (because Bohmian trajectories cannot cross). 

\item Absolute uncertainty sets a limit on our knowledge of a system's particle configuration in terms of its wave function. It does not say that our knowledge of a system is limited \emph{to} its wave function. 

Indeed, what we can know about wave functions is an entirely different question. It seems evident to me that our knowledge \emph{of} the wave function is usually much more limited -- and certainly much more indirect -- than our knowledge of particle positions. In fact, to the extent that we can measure the wave function (by so-called ``protective measurements'', see \cite{aharonov.vaidman1993}), we infer it from position measurements. 

\end{enumerate}

\noindent For all these reasons, attempts to use absolute uncertainty in an argument for the empirical priority of wave functions over particles are thoroughly misguided.

\section{Measurements and conscious experience}

All that said, some authors insist that Bohmian mechanics runs into problems when the description of the measurement process is supposed to end not with the pointer of a measurement device (or maybe a photograph of the measurement device) but the brain and conscious experience of an observer (e.g. \cite{gao2019a}, see \cite{oldofredi2019} for a good discussion). A priori, there are at least two reasons to be suspicious of such claims: \begin{enumerate}
\item Most of the authors making them seem to misunderstand Bohmian mechanics already as applied to measurement devices.

\item From the point of view of the physical theory, there is no essential difference between a measurement device and a brain (or whatever physical system is supposed to be the supervenience base of conscious experience). The particle configuration of a brain records an observation in the same sense as the particle configuration of a measurement device or a photographic film does. Everything else falls under the mind-body problem, about which, I believe, quantum physics has nothing new to say (cf. \cite{loewer2003}).
\end{enumerate}

\noindent Of course, there is in general more to a ``record'' than a static particle configuration. It is also relevant how the system in question evolves and interacts, and this is determined by the wave function. Thus, to the extent that there is a legitimate debate here, it comes down to the following question (cf. \cite{lewis2007}):
\begin{quote}If some functionalist theory of the mind is true, what makes it that mental states are functionally realized by the particles rather than the wave function which is also part of the Bohmian theory? \end{quote}
This objection is particularly popular among Everettians, who use it to argue that Bohmian mechanics is a Many-Worlds theory in denial (see, in particular, \cite{deutsch1996, brown.wallace2005}). Bohmian mechanics agrees, after all, that the wave function of the universe never collapses, thus admitting all the branches that make up the Everettian multiverse.

There are a few observations I can make in response:
\begin{enumerate}[a)]
\item I don't know if any functionalist theory of the mind is true (and I wouldn't want to make my understanding of quantum mechanics contingent on it).

\item To apply functionalism, it must be clear what the basic objects and properties are in terms of which the non-basic objects and properties are functionalized. In the Bohmian theory, the basic terms are particle positions, while the wave function is itself understood through its ``functional'' role for the motion of particles. 
 

\item There are good arguments for \emph{substrate independence} in the philosophy of mind, in particular the ``fading/dancing qualia'' of \cite{chalmers1995}, but I don't see how they would apply to particles and the wave function, even if the wave function were another physical entity (which I don't believe it is). It may be possible to gradually replace a biological brain by a silicone brain while maintaining the functional organization, but I don't know what it would even mean to replace parts of a particle brain by wave functions. 

\item In more detail, the objection against taking particles as the physical correlates of conscious experience is that ``If the functionalist assumption is correct, for consciousness to supervene on the Bohmian particles but not the wave function, the Bohmian particles must have some functional property that the wave function do not share. But the functional behaviour of the Bohmian particles is arguably identical to that of the branch of the wave function in which they reside.'' \cite[p. 306]{gao2019a}

The last assertion is also arguably false. For instance, Bohmian mechanics allows for the possibility that the universal wave function is stationary while all the change in the world comes from particle motions. Changing in time versus not changing in time is clearly a significant functional difference. I'm not committing to a stationary wave function, here; my point is that it cannot be \emph{a priori} true that anything which can be functionalized in terms of particles can also be functionalized in terms of the wave function, even under a generous interpretation of functionalism. 

\item On a more basic note: particles move relative to one another. The wave packets guiding their motion (to the extent that they are even separable) don't. They ``live'' in different dimensions of configuration space and hence do not even stand in a distance relation to one another.

Anything that allows for a functional definition in terms of matter in motion (this arguably includes brains, though the critical question is, of course, whether it includes ``minds'') can, in principle, be realized by particles. It is not clear at all that it can also be realized by degrees of freedom in the wave function.\footnote{What some neo-Everettians seem to establish is nothing more than a mapping between patterns in the wave function and trajectories in physical space. This is not even a mathematical isomorphism, let alone a functional one. I find it remarkable how the philosophical discussion turned to the question whether the empirical content of Bohmian mechanics is really that of Everettian quantum mechanics when it is not clear if Everettian quantum mechanics has any empirical content at all.}


\item Another version of the objection against particles as the physical correlates of conscious experience is that a conscious agent would then have precise knowledge of the particle configuration of her brain,  which leads to worries about faster-than-light signaling as well as to the question, how the brain measures it's own particle configuration \cite{stone1994}. To be honest, I don't even see how this objection gets off the ground. Knowledge realized in (or supervenient on) brain configurations is not knowledge \emph{about} brain configurations. 
\end{enumerate} 

\noindent In the upshot, to say that Bohmian mechanics cannot account for conscious experience (to the extent that it is physical) is to say that particles moving in accordance with the Bohmian laws cannot possibly be a ``brain''. As far as I can tell, this claim has no basis in physics, neuroscience, or anywhere else. On the other hand, the claim that ``brains'' would have to be located in the undulating wave function rather than moving particles is based on a variety of physical and metaphysical assumptions that are questionable, at best. I don't believe that physics can tell us why brain states are correlated with mental states but I believe that physics \emph{must} tell us what brains are made of. And the answer of Bohmian mechanics is clearly and unequivocally: particles.

\section{Epilogue}
When all is said and done, I suspect that some readers will still insist on the question:\\

\noindent \emph{Supppose that Bohmian mechanics is true, how do I know that the tree in front of me is a collection of particles rather than a pattern in the wave function?}\\

In response, I could insist on a particular metaphysical interpretation of the wave function and say that it is not physical stuff but rather a nomological object (see e.g. \cite{esfeld.etal2014}). I believe that this response is correct but doubt that it would satisfy the questioner. Thus, if I may be more blunt, I would say: if you even ask this question, you still have some physical theory in mind that is not Bohmian mechanics. I suppose that when you first studied classical Hamiltonian mechanics, you didn't wonder why, according to that theory, a tree is a configuration of particles rather than a pattern in the Hamiltonian flow on phase space.  Physics has never been about locating  trees in an abstract mathematical formalism, only the confusions about quantum mechanics lead to this business of ``interpretation''. Instead, the scientific enterprise departs from our ``manifest image of the world'' \citep{sellars1962}, our observation of trees, tables, cats, etc., and the question, what these objects are made of on the most fundamental level. Once we have a hypothesis about the basic entities and the laws describing them, we are in the business of locating trees (and cats, and measurement devices, etc.) in the scientific image of the theory to see if it matches the world that we experience. \\

\noindent\emph{But if the world -- including you -- was just patterns in the wave function rather than configurations of particles, your experience would be the same. }\\

I doubt that this is true, and the people who claim it is, have, again, another theory in mind than Bohmian mechanics. I agree that \emph{if} trees were patterns in the wave function, Bohmian mechanics would not be the correct theory of the world. However, what some physicists and philosophers have tried to argue is that even if Bohmian mechanics were true, the tree in front of you would most likely be a pattern in the wave function rather than a collection of particles. And these arguments don't hold water; they are question-begging at best and usually based on misconceptions of the physical theory. \\

\noindent\emph{I feel like you're still avoiding the real issue, so let me rephrase it: How does it follow FROM THE EQUATIONS of Bohmian mechanics that the tree in front of me is a configuration of particles rather than a pattern in the wave function? }\\

Nothing physical follows from mathematics alone. 
This is why the primitive ontology -- the stuff that trees are made of (or maybe instantiated in) -- is a basic and indispensable part of any fundamental physical theory. A theory with a clear primitive ontology can be wrong about what matters is, but it cannot be wrong about what it says that matter is. \\

\noindent \textbf{Acknowledgements:} I am grateful to Andrea Oldofredi for helpful comments and to Shan Gao for an inspiring discussion. I gratefully acknowledge funding by the Swiss National Science Foundation (SNSF) Doc.Mobility Fellowship P1LAP1\_184150.

\bibliographystyle{apalike}
\bibliography{Gao}

\begin{thebibliography}{}

\bibitem[Aharonov and Vaidman, 1993]{aharonov.vaidman1993}
Aharonov, Y. and Vaidman, L. (1993).
\newblock Measurement of the {{Schr\"odinger}} wave of a single particle.
\newblock {\em Physics Letters A}, 178(1):38--42.

\bibitem[Aharonov and Vaidman, 1996]{aharonov.vaidman1996}
Aharonov, Y. and Vaidman, L. (1996).
\newblock About {{Position Measurements Which}} do {{Not Show}} the {{Bohmian
  Particle Position}}.
\newblock In Cushing, J.~T., Fine, A., and Goldstein, S., editors, {\em Bohmian
  {{Mechanics}} and {{Quantum Theory}}: {{An Appraisal}}}, Boston {{Studies}}
  in the {{Philosophy}} of {{Science}}, pages 141--154. {Springer Netherlands},
  Dordrecht.

\bibitem[Allori et~al., 2014]{allori.etal2014}
Allori, V., Goldstein, S., Tumulka, R., and Zangh\`i, N. (2014).
\newblock Predictions and primitive ontology in quantum foundations: A study of
  examples.
\newblock {\em British Journal for the Philosophy of Science}, 65(2):323--352.

\bibitem[Bedard, 1999]{bedard1999}
Bedard, K. (1999).
\newblock Material {{Objects}} in {{Bohm}}'s {{Interpretation}}.
\newblock {\em Philosophy of Science}, 66(2):221--242.

\bibitem[Bell, 2004]{bell2004}
Bell, J.~S. (2004).
\newblock {\em Speakable and Unspeakable in Quantum Mechanics}.
\newblock {Cambridge: Cambridge University Press}, second edition.

\bibitem[Bohm, 1952]{bohm1952a}
Bohm, D. (1952).
\newblock A suggested interpretation of the quantum theory in terms of
  ``hidden'' variables. 2.
\newblock {\em Physical Review}, 85(2):180--193.

\bibitem[Brown and Wallace, 2005]{brown.wallace2005}
Brown, H. and Wallace, D. (2005).
\newblock Solving the {{Measurement Problem}}: {{De Broglie}}--{{Bohm Loses
  Out}} to {{Everett}}.
\newblock {\em Foundations of Physics}, 35(4):517--540.

\bibitem[Chalmers, 1995]{chalmers1995}
Chalmers, D.~J. (1995).
\newblock Absent {{Qualia}}, {{Fading Qualia}}, {{Dancing Qualia}}.
\newblock In Metzinger, T., editor, {\em Conscious {{Experience}}}, pages
  309--328. {Ferdinand Schoningh}.

\bibitem[Deutsch, 1996]{deutsch1996}
Deutsch, D. (1996).
\newblock Comment on {{Lockwood}}.
\newblock {\em The British Journal for the Philosophy of Science},
  47(2):222--228.

\bibitem[D\"urr et~al., 2004]{durr.etal2004}
D\"urr, D., Goldstein, S., and Zangh\`i, N. (2004).
\newblock Quantum {{Equilibrium}} and the {{Role}} of {{Operators}} as
  {{Observables}} in {{Quantum Theory}}.
\newblock {\em Journal of Statistical Physics}, 116(1):959--1055.
\newblock Reprinted in D\"urr et al. (2013, ch. 3).

\bibitem[D\"urr et~al., 2013]{durr.etal2013a}
D\"urr, D., Goldstein, S., and Zangh\`i, N. (2013).
\newblock {\em Quantum Physics without Quantum Philosophy}.
\newblock {Berlin: Springer}.

\bibitem[Englert et~al., 2014]{englert.etal2014}
Englert, B.-G., Scully, M.~O., S\"ussmann, G., and Walther, H. (2014).
\newblock Surrealistic {{Bohm Trajectories}}.
\newblock {\em Zeitschrift f\"ur Naturforschung A}, 47(12):1175--1186.

\bibitem[Esfeld et~al., 2014]{esfeld.etal2014}
Esfeld, M., Lazarovici, D., Hubert, M., and D\"urr, D. (2014).
\newblock The ontology of {{Bohmian}} mechanics.
\newblock {\em British Journal for the Philosophy of Science}, 65(4):773--796.

\bibitem[Gao, 2019a]{gao2019}
Gao, S. (2019a).
\newblock A contradiction in {{Bohm}}'s theory.
\newblock Preprint: http://philsci-archive.pitt.edu/15713/.

\bibitem[Gao, 2019b]{gao2019a}
Gao, S. (2019b).
\newblock The measurement problem revisited.
\newblock {\em Synthese}, 196(1):299--311.

\bibitem[Gisin, 2018]{gisin2018}
Gisin, N. (2018).
\newblock Why {{Bohmian Mechanics}}? {{One}}- and {{Two}}-{{Time Position
  Measurements}}, {{Bell Inequalities}}, {{Philosophy}}, and {{Physics}}.
\newblock {\em Entropy}, 20(2):105.

\bibitem[Heisenberg, 2012]{heisenberg2012}
Heisenberg, W. (2012).
\newblock {\em {Der Teil und das Ganze: Gespr\"ache im Umkreis der
  Atomphysik}}.
\newblock {Piper Verlag}, 9 edition.

\bibitem[Lazarovici et~al., 2018]{lazarovici.etal2018}
Lazarovici, D., Oldofredi, A., and Esfeld, M. (2018).
\newblock Observables and {{Unobservables}} in {{Quantum Mechanics}}: {{How}}
  the {{No}}-{{Hidden}}-{{Variables Theorems Support}} the {{Bohmian Particle
  Ontology}}.
\newblock {\em Entropy}, 20(5):381.

\bibitem[Lewis, 2007]{lewis2007}
Lewis, P.~J. (2007).
\newblock How {{Bohm}}'s {{Theory Solves}} the {{Measurement Problem}}.
\newblock {\em Philosophy of Science}, 74(5):749--760.

\bibitem[Loewer, 2003]{loewer2003}
Loewer, B.~M. (2003).
\newblock Consciousness and {{Quantum Theory}}: {{Strange Bedfellows}}.
\newblock In Smith, Q. and Jokic, A., editors, {\em Consciousness: {{New
  Philosophical Perspectives}}}. {Oxford University Press}.

\bibitem[Maudlin, 1995a]{maudlin1995e}
Maudlin, T. (1995a).
\newblock Three measurement problems.
\newblock {\em Topoi}, 14(1):7--15.

\bibitem[Maudlin, 1995b]{maudlin1995d}
Maudlin, T. (1995b).
\newblock Why {{Bohm}}'s {{Theory Solves}} the {{Measurement Problem}}.
\newblock {\em Philosophy of Science}, 62(3):479--483.

\bibitem[Maudlin, 2010]{maudlin2010}
Maudlin, T. (2010).
\newblock Can the world be only wave-function?
\newblock In Saunders, S., Barrett, J., Kent, A., and Wallace, D., editors,
  {\em Many Worlds? {{Everett}}, Quantum Theory, and Reality}, pages 121--143.
  {Oxford: Oxford University Press}.

\bibitem[Monton, 2006]{monton2006}
Monton, B. (2006).
\newblock Quantum mechanics and {{3N}}-dimensional space.
\newblock {\em Philosophy of science}, 73(5):778--789.

\bibitem[{Naaman-Marom} et~al., 2012]{naaman-marom.etal2012}
{Naaman-Marom}, G., Erez, N., and Vaidman, L. (2012).
\newblock Position measurements in the de {{Broglie}}\textendash{{Bohm}}
  interpretation of quantum mechanics.
\newblock {\em Annals of Physics}, 327(10):2522--2542.

\bibitem[Oldofredi, 2019]{oldofredi2019}
Oldofredi, A. (2019).
\newblock Some remarks on the mentalistic reformulation of the measurement
  problem: A reply to {{S}}. {{Gao}}.
\newblock {\em Synthese}.

\bibitem[Sellars, 1962]{sellars1962}
Sellars, W. (1962).
\newblock Philosophy and the scientific image of man.
\newblock In Colodny, R., editor, {\em Frontiers of Science and Philosophy},
  pages 35--78. {Pittsburgh: University of Pittsburgh Press}.

\bibitem[Stone, 1994]{stone1994}
Stone, A.~D. (1994).
\newblock Does the {{Bohm Theory Solve}} the {{Measurement Problem}}?
\newblock {\em Philosophy of Science}, 61(2):250--266.

\bibitem[Zeh, 1999]{zeh1999}
Zeh, H.~D. (1999).
\newblock Why {{Bohm}}'s {{Quantum Theory}}?
\newblock {\em Foundations of Physics Letters}, 12(2):197--200.

\end{thebibliography}

\end{document}